%% file: GTRG.tex
\documentclass[a4paper,11pt]{article}
\usepackage{jheppub} 
\usepackage{graphicx}
\usepackage{hyperref}
\usepackage{natbib}

\newcommand{\diff}{\mathrm{d}}

\newcommand{\e}{\mathrm{e}}

\newcommand{\gTr}{\mathrm{gTr}}
\newcommand{\tTr}{\mathrm{tTr}}
\title{More about the Grassmann tensor renormalization group}

 \author[a]{Shinichiro Akiyama,}
 \emailAdd{akiyama@het.ph.tsukuba.ac.jp}

 \author[b,c]{Daisuke Kadoh}
 \emailAdd{kadoh@keio.jp}

 \affiliation[a]{Graduate School of Pure and Applied Sciences, University of Tsukuba, Tsukuba, Ibaraki
    305-8571, Japan}
 \affiliation[b]{Physics Division, National Center for Theoretical Sciences, National Tsing-Hua University,\\ Hsinchu, 30013, Taiwan}
 \affiliation[c]{Research and Educational Center for Natural Sciences, Keio University, \\ Yokohama 223-8521, Japan}

\abstract{
We derive a general formula of the tensor network representation for $d$-dimensional lattice fermions
with ultra-local interactions, including Wilson fermions, staggered fermions, and domain-wall fermions. 
The Grassmann tensor is concretely defined with auxiliary Grassmann variables that play a role in bond degrees of freedom.
Compared to previous works, our formula does not refer to the details of lattice fermions and is
derived by using the singular value decomposition for the given Dirac matrix without
any ad-hoc treatment for each fermion. 
We numerically test our formula for free Wilson and staggered fermions and find that it properly works for them. We also find that Wilson fermions show better performance than staggered fermions in the tensor renormalization group approach, unlike the Monte Carlo method.
}

\begin{document}
\preprint{UTHEP-751}
\maketitle
\flushbottom

\section{Introduction}
\label{sec:introduction}

Tensor renormalization group (TRG) is a promising computational approach to study the lattice field theory. The dynamics of the theory can be investigated by the TRG mostly in the thermodynamic limit without suffering from the sign problem since it does not employ any stochastic process. The TRG was originally proposed by Levin and Nave as a real space renormalization group for the two-dimensional Ising model \cite{Levin:2006jai}. Extensions to fermionic systems were firstly discussed by Gu {\it et al.} \cite{Gu:2010yh,Gu:2013gba},  and the Grassmann TRG has been applied to many models such as the Schwinger model with and without $\theta$ term \cite{Shimizu:2014uva,Shimizu:2014fsa,Shimizu:2017onf}, the Gross-Neveu model with finite density \cite{Takeda:2014vwa}, and $\mathcal{N}=1$ Wess-Zumino model \cite{Kadoh:2018hqq}.
\footnote{See Refs.~\cite{Sakai:2017jwp,Yoshimura:2017jpk,Meurice:2018fky,PhysRevD.101.094509} for other related studies. } These earlier works have verified that the TRG is also useful to evaluate 
the path integral over the Grassmann variables.

In this paper, we introduce a different way from these studies to derive a tensor network representation
and to implement the Grassmann TRG. 
The initial tensor network and renormalized ones are treated in a unified manner. 
We derive a general tensor network formula, characterized by the singular value decomposition (SVD) of a given Dirac matrix, 
for any local lattice fermion such as Wilson fermions, staggered fermions, or domain-wall fermions.  
Our method has several advantages over previous methods.
As seen later,  it is useful in comparing tensor networks for various lattice fermions from common aspects (for instance, computational times and errors). 
In previous works, the subscript structure of initial tensors is different from those of renormalized tensors. \footnote{See, for instance, Eqs.~(4) and (36) of Ref.~\cite{Sakai:2017jwp}. }
Such an inhomogeneity of tensors requires separate handling of tensors for the initial and the renormalized steps. 
That kind of extra handling is absent in our method. 
With our method, 
several proposed TRG methods applicable for the Ising model are naturally transcribed to ones with fermions.

This paper is organized as follows. In Sec. \ref{sec:formulation}, we define a Grassmann tensor and its contraction rule. We explain how to construct the tensor network with auxiliary Grassmann fields in Sec. \ref{sec:TN}. 
To test our method, 
numerical results for the two-dimensional free Wilson and staggered fermions
are provided in Sec. \ref{sec:GTRG}. Sec. \ref{sec:summary} is 
devoted to summary and outlook.
Truncation technique is explained in Appendix \ref{sec:truncation}. 
Concrete forms of Grassmann tensor for 2D Wilson and staggered fermions are shown in Appendix \ref{sec:2DTN}.

\section{Formalism of the Grassmann tensors}
\label{sec:formulation}

The variables $\eta_i \ (i=1,\cdots,N) $ are single-component Grassmann numbers which satisfy the anti-commutation relation $ \{\eta_{i},\eta_{j} \}=0 $. We begin with defining a Grassmann tensor and its contraction rule with single-component index $\eta_i$. 
\footnote{
The Grassmann contraction is also discussed in Refs.~\cite{Gu:2010yh,Gu:2013gba} and the appendix B of Ref.~\cite{Bao:2019hfc}.
Eqs.~\eqref{eq:def1} and \eqref{eq:oriented} defined below generalize the results of section $3.3$ of Ref.~\cite{Meurice:2018fky}.
}
Then those with multi-component index $\Psi=(\eta_1,\eta_2,\cdots,\eta_N)$ are defined by extending the single component case straightforwardly.

\begin{figure}[htbp]
	\centering
	\includegraphics[width=0.55\hsize,bb= 0 0 792 612]{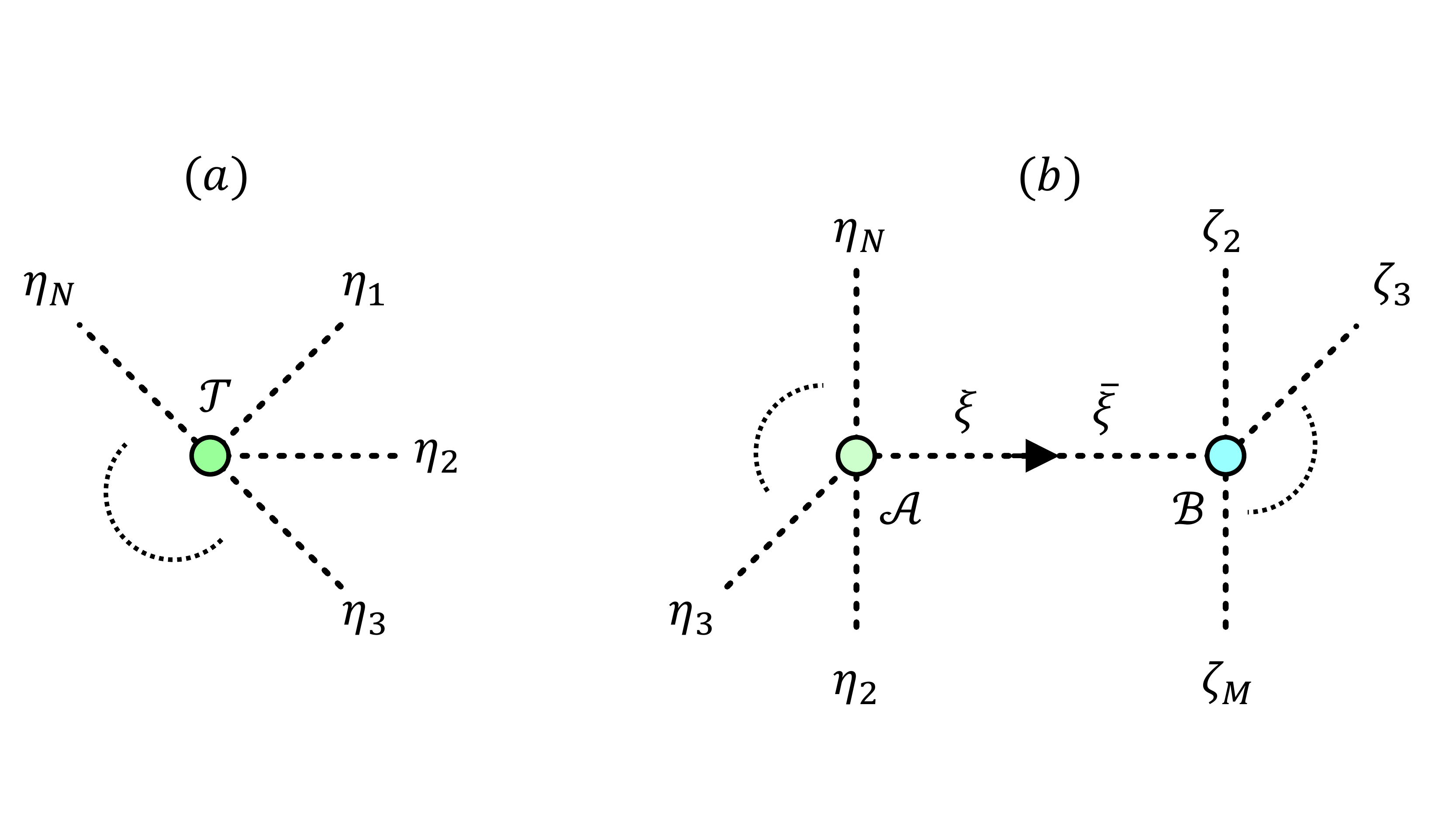}
	\caption{Graphical representations of (a) Grassmann tensor in Eq.\eqref{eq:def1} and  (b) Grassmann contraction in Eq.~\eqref{eq:oriented}. The external lines specify uncontracted indices.  
The arrow in the internal line represents a contracted direction from $\xi$ to $\bar \xi$. 
}
	\label{fig:fig1}
\end{figure}

The {\it Grassmann tensor} $\mathcal{T}$ of rank $N$ is defined as
\begin{align}
\label{eq:def1}
	\mathcal{T}_{\eta_{1}\eta_{2}\cdots\eta_{N}}=\sum_{i_{1}=0}^1 \sum_{i_{2}=0}^1 \cdots \sum_{i_{N}=0}^1 T_{i_{1}i_{2}\cdots i_{N}}\eta_{1}^{i_{1}}\eta_{2}^{i_{2}}\cdots\eta_{N}^{i_{N}},
\end{align}
where $\eta_i$ are single-component Grassmann numbers and $T_{i_{1}i_{2}\cdots i_{N}}$ is referred to as a coefficient tensor, whose rank is also $N$, with complex entries. Fig.~\ref{fig:fig1}~(a) represents a Grassmann tensor, where the external lines correspond to the indices $\eta_i$.

We consider a {\it Grassmann contraction} among Grassmann tensors. 
Let $\mathcal{A}_{\eta_1\ldots\eta_N}$ and $\mathcal{B}_{\zeta_1\ldots\zeta_M}$ be two Grassmann tensors of rank $N$ and $M$, respectively.
\footnote{ We assume that either of $\mathcal{A}$ and $\mathcal{B}$ is a commutative tensor whose coefficient tensor $A(B)_{i_1i_2 \cdots i_{N(M)}}=0$ for $(i_1+i_2 +\cdots i_{N(M)}) \, {\rm mod} \, 2 =1$. In this case, $\mathcal{A}_{\eta_1\ldots\eta_N} \mathcal{B}_{\zeta_1\ldots\zeta_M} =\mathcal{B}_{\zeta_1\ldots\zeta_M} \mathcal{A}_{\eta_1\ldots\eta_N}$ and Eq.~\eqref{eq:oriented} can also be expressed as $\int\diff\bar{\xi}\diff\xi~\e^{-\bar{\xi}\xi}\mathcal{B}_{\bar{\xi}\zeta_2 \ldots \zeta_M}\mathcal{A}_{\xi\eta_2\ldots \eta_N}$.
}
The Grassmann contraction has an orientation which comes from the anti-commutation relation of 
Grassmann variables. We define a Grassmann contraction from $\eta_1$ to $\zeta_1$ as
\begin{align}
\label{eq:oriented}
	\int\diff\bar{\xi}\diff\xi~\e^{-\bar{\xi}\xi}
\mathcal{A}_{\xi\eta_2\ldots \eta_N}\mathcal{B}_{\bar{\xi}\zeta_2 \ldots \zeta_M}.
\end{align}
Eq.~\eqref{eq:oriented} itself is a Grassmann tensor, and the coefficient tensor of Eq.\eqref{eq:oriented} is given by a contraction of two coefficient tensors  of $\mathcal{A}$ and $\mathcal{B}$ with some sign factors. 
One can consider a contraction from $\eta_i$ to $\zeta_j$ 
as a straightforward extension of Eq.\eqref{eq:oriented} with keeping the weight factor $\e^{-\bar{\xi}\xi}$. In Fig.~\ref{fig:fig1}~(b), the Grassmann contraction is shown as a shared link with the arrow.
Note that $\int\diff\bar{\xi}\diff\xi \, \e^{-\bar{\xi}\xi}\mathcal{A}_{\bar\xi\eta_2\ldots \eta_N}\mathcal{B}_{{\xi}\zeta_2 \ldots \zeta_M}$ should be represented as  Fig.~\ref{fig:fig1}~(b) with the opposite arrow.

\begin{figure}[htbp]
	\centering
	\includegraphics[width=0.55\hsize,bb= 0 0 792 612]{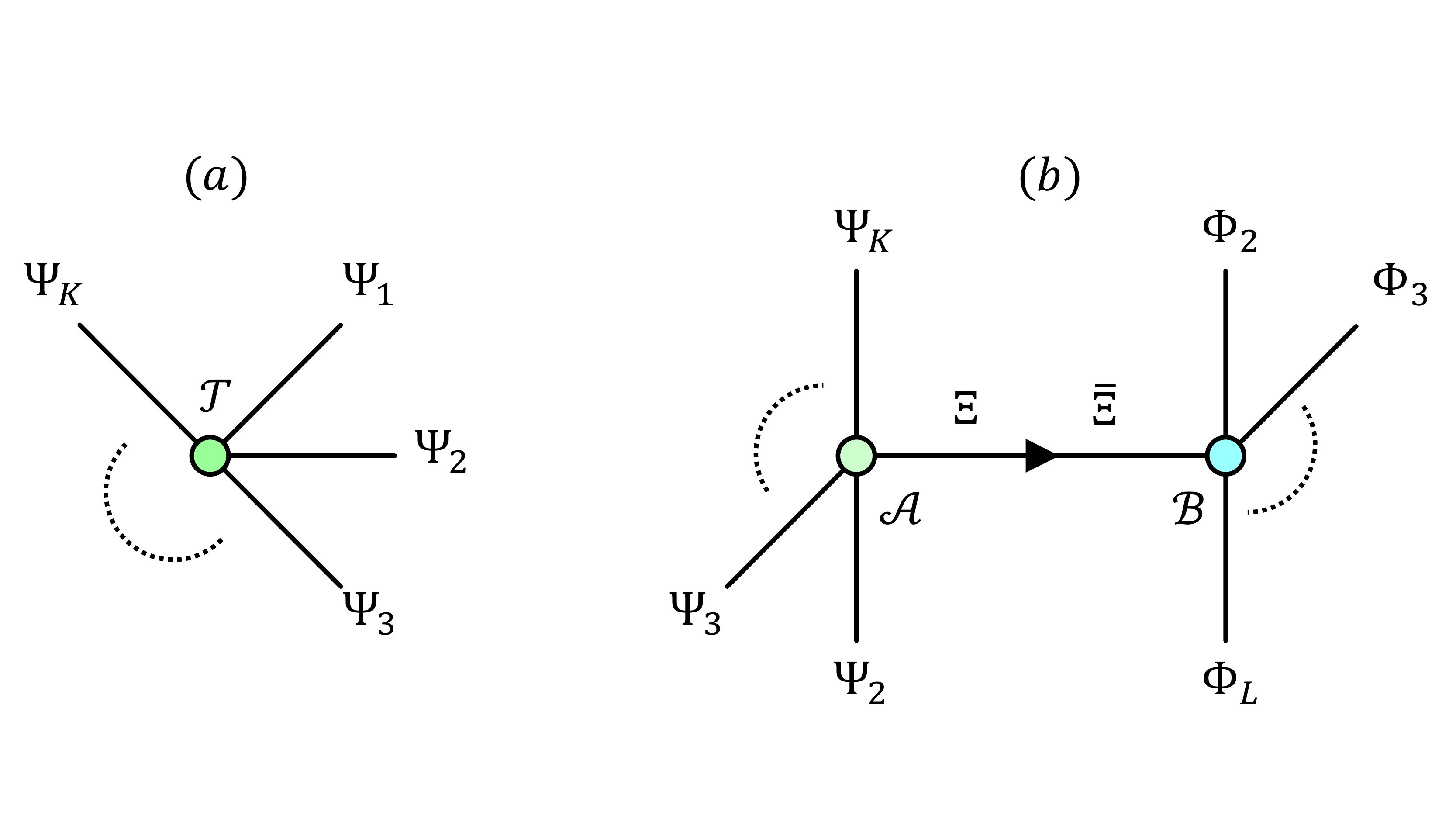}
\caption{Graphical representations of (a) Grassmann tensor in Eq.\eqref{eq:def_multi} and  (b) Grassmann contraction in Eq.~\eqref{eq:oriented_multi}. The external lines specify uncontracted multi-component indices. The arrow in the internal line represents a contracted direction from $\Xi$ to $\bar \Xi$. 
}
	\label{fig:fig1m}
\end{figure}

Let us now move on to the multi-component case. For simplicity of explanation, we take $N=mK$ for Eq.~\eqref{eq:def1}. Then $N$ Grassmann numbers $\eta_n$ are divided into $m$ component variables $\Psi_a$ ($a=1,\cdots,K$) as $\Psi_a=(\eta_{(a-1)m+1}, \eta_{(a-1)m+2} ,\cdots,\eta_{am})$. The Grassmann tensor Eq.~\eqref{eq:def1} is also expressed as 
\begin{align}
\label{eq:def_multi}
	\mathcal{T}_{\Psi_1 \Psi_2 \cdots \Psi_K}  \equiv  \mathcal{T}_{\eta_{1}\eta_{2}\cdots\eta_{N}}
\end{align}
The rank of a Grassmann tensor should be carefully read from the dimension of indices since both sides of Eq.~\eqref{eq:def_multi} have the same rank. Fig.~\ref{fig:fig1m}~(a) represents an example of Grassmann tensor with multi-component indices, where the external links are shown as solid lines correspond to the multi-component indices $\Psi_a$. Other cases are straightforwardly generalized from Eq.\eqref{eq:def_multi}.

To define the Grassmann contraction with multi-dimensional indices, we consider the case of $N=mK, M=mL$ in Eq.\eqref{eq:oriented} for simplicity. The $N$ and $M$ rank tensors $\mathcal{A}_{\eta_1\eta_2\cdots\eta_{N}}$ and $\mathcal{B}_{\zeta_1 \zeta_2\cdots \zeta_{M}}$ are expressed as $\mathcal{A}_{\Psi_1 \Psi_2 \cdots \Psi_K}$ and $\mathcal{B}_{\Phi_1 \Phi_2 \cdots \Phi_L}$ 
where $\Psi_a$ and $\Phi_a$ are $m$-component indices defined as in Eq.~\eqref{eq:def_multi}. Then the Grassmann contraction is given for the multi-component case: 
\begin{align}
\label{eq:oriented_multi}
	\int\diff\bar{\Xi}\diff\Xi \, 
\mathcal{A}_{\Xi\Psi_2\ldots \Psi_K}\mathcal{B}_{\bar{\Xi}\Phi_2 \ldots \Phi_L}
\end{align}
where $\Xi=(\xi_1,\xi_2,\cdots,\xi_m), \bar \Xi=(\bar \xi_m,\cdots,\bar \xi_2, \bar \xi_1)$ and 
\begin{align}
      \label{eq:oriented_multi_2}
	\diff\bar{\Xi}\diff\Xi 
	\equiv \prod_{n=1}^{m}\diff\bar{\xi}_{n}\diff\xi_{n}~\e^{-\bar{\xi}_{n}\xi_{n}}.
\end{align}
The case of $m=1$ reproduces Eq.~\eqref{eq:oriented}. We should note that $\bar \Xi$ contains $\bar\xi_n$ in a reverse order so that the coefficient tensor of Eq.\eqref{eq:oriented_multi} is simply given by a tensor contraction of coefficient tensors of $\mathcal{A}$ and $\mathcal{B}$ without extra sign factors. Fig.~\ref{fig:fig1m}~(b) shows the Grassmann contraction with multi-component indices.

It is easy to define a {\it Grassmann tensor network} with these notations. Let $\mathcal{T}_n$ be Grassmann tensors. Then the tensor network is defined by a product of $\mathcal{T}_1\mathcal{T}_2\cdots$ where all indices are contracted as  Eq.~\eqref{eq:oriented_multi}.

\section{General formula of tensor network for arbitrary lattice fermions}
\label{sec:TN}

We prove that the path integral of lattice fermion theory with nearest-neighbor interactions is expressed as a Grassmann tensor network.  We assume that the theory has translational invariance on the lattice.
The $d$-dimensional hypercubic lattice is defined by a set of integer lattice sites $\Lambda  =\{(n_1,n_2,\cdots,n_d) \, | \ n_i \in \mathbb{Z} \ {\rm for} \, i=1,2,\cdots,d \}$ where the lattice spacing $a$ is set to $a=1$. 
Although we begin with a lattice action defined by quadratic forms of Grassmann variables, 
it is straightforward to include four-fermion interactions. 
Next-nearest-neighbor and higher interactions are also easily included because these terms are expressed as nearest-neighbor ones using auxiliary Grassmann variables. 

Consider lattice  fermion fields $\psi_a(n)$ and $\bar\psi_{a}(n)$ for $n \in \Lambda$ where  $a$ runs from $1$ to $N$, which is the degree of freedom of the internal space such as the spinor or the flavor space. Then the lattice fermion action is formally given by
\begin{align}
	S= \sum_{n\in \Lambda} \bar\psi(n) (D \psi)(n)
\end{align}
where $D$ is the Dirac operator acting on the fermion field as
\begin{align}
	(D \psi)_a(n) = \sum_{m \in \Lambda} \sum_{b=1}^N D_{ab}(n,m) \psi_b(m). 
\end{align}
We may consider that $D$ takes a form of 
\begin{align}
	D_{ab}(n,m) = W_{ab} \delta(n,m) + \sum_{\mu=1}^d (X_\mu)_{ab} \delta(n+\hat \mu,m) 
	+ \sum_{\mu=1}^d  (Y_\mu)_{ab} \delta(n-\hat \mu,m)
	\label{gerenal_form}
\end{align}
without loss of generality. Here, 
$\delta(n,m)$ is the Kronecker delta. 
$X_\mu,Y_\mu,W$ are matrices with respect to the internal space. 
The $W$ term is an on-site interaction and the $X_\mu$ and $Y_\mu$ terms are nearest-neighbor interactions. The path integral is defined as
\begin{align}
Z= \int \left[{\rm D}\psi {\rm D}\bar\psi\right] \, \e^{-S} 
\label{Z_general}
\end{align}
where 
$[{\rm D}\psi {\rm D}\bar\psi] = \prod_{n \in \Lambda}  \prod_{a=1}^N \diff\psi_a(n) \diff \bar\psi_a(n)$ 
with single-component Grassmann measures  $\diff \psi_a(n)$ and $\diff \bar\psi_a(n)$.

Let us firstly consider $X_\mu$ term in the action, dropping the spacetime index $n, \mu$ in the following for simplicity. The SVD of $X_{ab}$ is given by $X_{ab}=\sum_{c=1}^N U_{ac} \sigma_c (V^\dag)_{cb}$ where $\sigma_c \ge0$ are singular values and $U,V$ are unitary matrices. Then we have
\begin{align}
\label{eq:step1}
	\bar{\psi} X \psi= \sum_{c=1}^N \sigma_{c}\bar{\chi}_c \chi_c
\end{align}
where $\bar\chi = \bar\psi U$ and  $\chi = V^\dag\psi$. See Refs.~\cite{Shimizu:2014uva,Shimizu:2014fsa,Shimizu:2017onf,Takeda:2014vwa,Kadoh:2018hqq} 
and the discussion in Ref.~\cite{Meurice:2012wp} for the similar deformation. 
Using an identity,  
\begin{align}
\label{eq:identity}
	\e^{- \sigma_{c} \bar{\chi}_c  \chi_c}
	=\int\diff\bar{\eta}_c \diff\eta_c
	\exp\left[-\bar{\eta}_c\eta_c -\bar{\chi}_c {\eta}_c +  \sigma_{c}  \bar\eta_c \chi_c \right],
\end{align}
we can easily show that
\begin{align}
\label{eq:basic_id1}
	&\e^{-\bar{\psi}(n) X_\mu \psi(n+\hat\mu) } %\nonumber\\
	=  \prod_{c=1}^{K_\mu}  
	\int\diff\bar{\eta}_{\mu,c}(n) \diff\eta_{\mu,c}(n)~{\rm e}^{-\bar{\eta}_{\mu,c}(n) \eta_{\mu,c}(n)}\nonumber\\
	& \hspace{2.5cm} \times\exp\left[ -\{\bar{\psi}(n)U_{X_\mu}\}_{c}  \eta_{\mu,c}(n) +(\sigma_{X_\mu})_{c}  \bar{\eta}_{\mu,c}(n) \{V^{\dag}_{X_\mu}\psi(n+\hat\mu)\}_{c} \right],
\end{align}
where $\sigma_{X_\mu}$  and $U_{X_\mu}, V_{X_\mu}$ are singular values and singular vectors of $X_\mu$. Here $n,\mu$ dependences are explicitly shown. Similarly, 
\begin{align}
\label{eq:basic_id2}
	&\e^{-\bar{\psi}(n+\mu) Y_\mu \psi(n) } %\nonumber\\
	=  \prod_{c=1}^{L_\mu}  
	\int\diff\bar{\zeta}_{\mu,c}(n) \diff\zeta_{\mu,c}(n)~{\rm e}^{-\bar{\zeta}_{\mu,c}(n) \zeta_{\mu,c}(n)}\nonumber\\
	&\hspace{3cm} \times\exp\left[\{\bar{\psi}(n+\mu)U_{Y_\mu}\}_{c}  \bar{\zeta}_{\mu,c}(n) + (\sigma_{Y_\mu})_{c}  \zeta_{\mu,c}(n)\{V^{\dag}_{Y_\mu}\psi(n)\}_{c} \right],
\end{align}
where $\sigma_{Y_\mu}$  and $U_{Y_\mu}, V_{Y_\mu}$  are singular values and singular vectors of $Y_\mu$.

\begin{figure}[htbp]
	\centering
	\includegraphics[width=0.55\hsize,bb= 0 0 792 612]{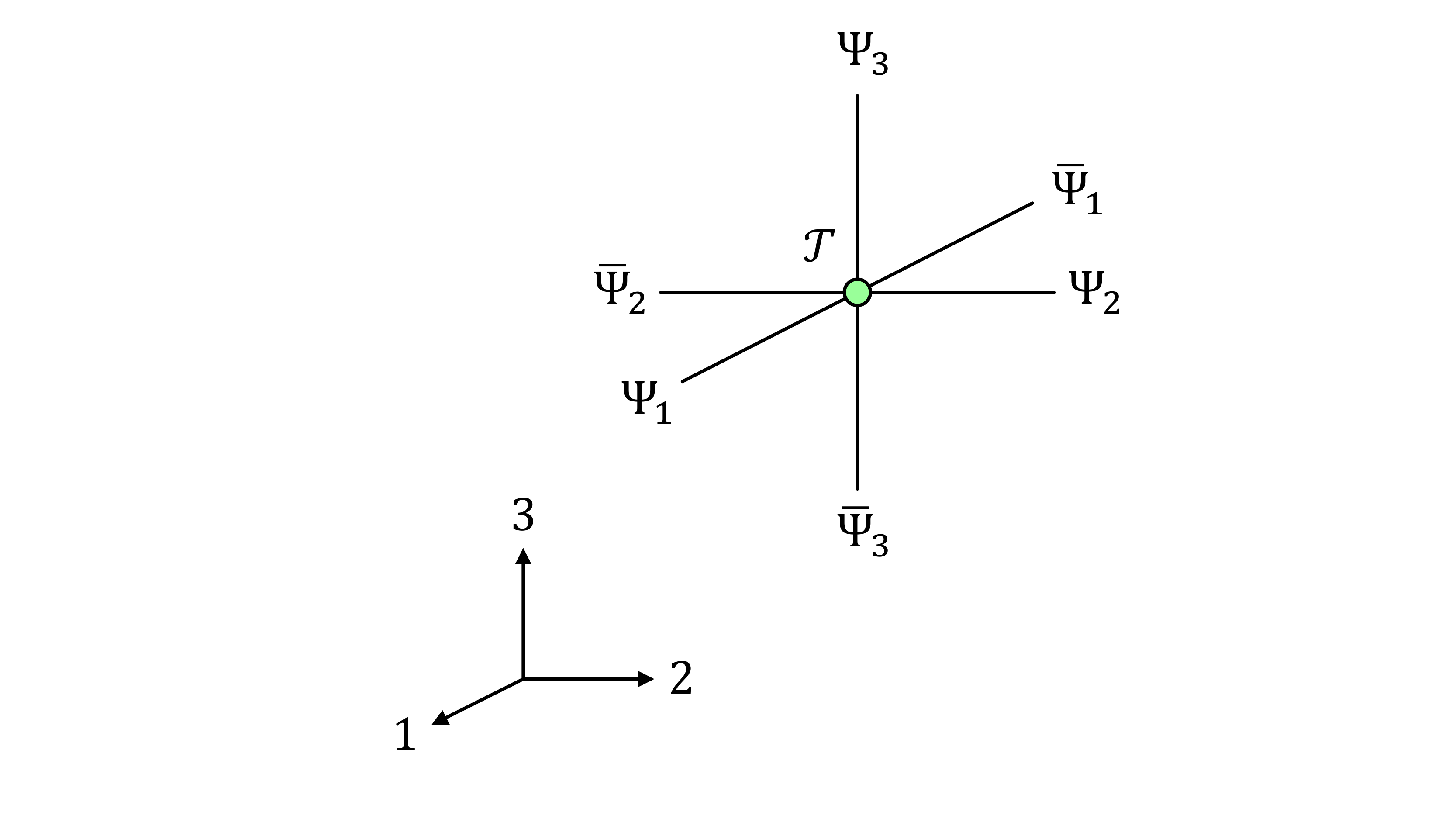}
	\caption{Graphical representation of Eq.~\eqref{eq:Gtensor} in three dimensions.}
	\label{fig:tensor}
\end{figure}

Inserting Eqs. (\ref{eq:basic_id1}) and (\ref{eq:basic_id2}) into Eq. (\ref{Z_general}) with Eq. (\ref{gerenal_form}),
we have
\begin{eqnarray}
\label{eq:Gtensor_rep}
	Z= \int \diff \bar \Psi \diff \Psi 
	\prod_{n \in \Lambda} 
	\mathcal{T}_{\Psi_1(n) \cdots \Psi_d(n)  \bar \Psi_d(n-\hat d) \cdots \bar\Psi_1 (n-\hat 1)}
\end{eqnarray}
where 
\begin{align}
\label{eq:Gtensor}
	&\mathcal{T}_{\Psi_1(n) \cdots \Psi_d(n)  \bar \Psi_d(n-\hat d) \cdots \bar\Psi_1 (n-\hat 1)} 
	=
	\int \left(\prod_{a=1}^{N}\diff\psi_{a}\diff\bar{\psi}_{a} \right)
	\exp\left[-\bar{\psi} W \psi \right]\nonumber\\
	&\hspace{2cm} \times\exp\left[\sum_{\mu=1}^{d}
	\sum_{c=1}^{K_{\mu}}\left\{-\{\bar{\psi} U_{X_\mu}\}_{c}  \eta_{\mu,c}(n) 
	+ (\sigma_{X_\mu})_{c}  \bar{\eta}_{\mu,c}(n-\hat{\mu}) \{V_{X_{\mu}}^{\dag}\psi\}_{c}\right\}\right]\nonumber\\
	&\hspace{2cm} \times\exp\left[
	\sum_{\mu=1}^{d}\sum_{c=1}^{L_{\mu}}\left\{\{\bar{\psi}U_{Y_\mu}\}_{c}  \bar{\zeta}_{\mu,c}(n-\hat{\mu}) 
	+ (\sigma_{Y_\mu})_{c}  \zeta_{\mu,c}(n) \{V_{Y_{\mu}}^{\dag}\psi\}_{c}\right\}\right],
\end{align}
and
\begin{align}
& \diff \bar \Psi \diff \Psi  \equiv 
\prod_{n \in \Lambda} 
\prod_{\mu=1}^d 
\left( \prod_{c=1}^{K_\mu}  
\diff\bar{\eta}_{\mu,c}(n) \diff\eta_{\mu,c}(n)~ {\rm e}^{-\bar{\eta}_{\mu,c}(n) \eta_{\mu,c}(n)} \right) \nonumber \\
&\hspace{2.5cm} 
\times \left(
\prod_{c=1}^{L_\mu}  \diff\bar{\zeta}_{\mu,c}(n) \diff\zeta_{\mu,c}(n)~{\rm e}^{-\bar{\zeta}_{\mu,c}(n) \zeta_{\mu,c}(n)} \right)
\end{align}
with $\Psi_\mu=(\eta_{\mu,1},\cdots,\eta_{\mu,K_{\mu}},\zeta_{\mu,1},\cdots,\zeta_{\mu,L_{\mu}})$ and $\bar\Psi_\mu=(\bar\zeta_{\mu,L_{\mu}},\cdots,\bar\zeta_{\mu,1},\bar\eta_{\mu,K_\mu},\cdots,\bar\eta_{\mu,1})$. Note that $\mathcal{T}$ is uniformly defined for the spacetime. 
It is easy to show Eq.~\eqref{eq:Gtensor_rep} by inserting Eq.~\eqref{eq:Gtensor} into it with 
identities Eq.~\eqref{eq:basic_id1} and Eq.~\eqref{eq:basic_id2}. Fig.~\ref{fig:tensor} shows Eq.~\eqref{eq:Gtensor} in three dimensions. Eq.~\eqref{eq:Gtensor_rep} is a Grassmann tensor network since a pair of $\Psi(n)$ and $\bar \Psi(n)$ appears once in Eq.~\eqref{eq:Gtensor_rep} under $\prod_{n \in \Lambda}$ and they are contracted with appropriate weights.

We denote Eq.~(\ref{eq:Gtensor_rep}) as 
\begin{align}
\label{eq:tn}
	Z=\gTr\left[\prod_{n\in\Lambda}\mathcal{T}_{\Psi_1(n) \cdots \Psi_d(n)  \bar \Psi_d(n-\hat d) \cdots \bar\Psi_1 (n-\hat 1)}\right]
\end{align}
where $\gTr$ means all possible Grassmann contractions defined in 
Eqs.~(\ref{eq:oriented_multi}) and (\ref{eq:oriented_multi_2}).
The situation is quite similar with the tensor network representation for spin models, which is denoted by $\tTr$ over tensor contractions on lattice.

This tensor network formulation is immediately applicable to any model with lattice fermions. It is also straightforward to extend the formulation to the models with next-nearest-neighbor and higher interactions because Eq.~\eqref{eq:identity} allows us to express a next-nearest-neighbor term as nearest-neighbor ones. Other on-site terms such as four-fermion interactions can also be included in Eq.~\eqref{eq:Gtensor} with no difficulty.

\section{Numerical applications}
\label{sec:GTRG}

The current Grassmann tensor network can be evaluated by coarse-graining algorithms 
with a truncation of degrees of freedom, 
such as the original Levin-Nave TRG \cite{Levin:2006jai} 
and some variations of the TRG \cite{PhysRevB.86.045139,Adachi:2019paf,Lan:2019stu,Kadoh:2019kqk}. 
In this section, we consider the higher-order TRG (HOTRG) \cite{PhysRevB.86.045139}, which is applicable to any dimensional lattices 
for a Grassmann tensor network.

\begin{figure}[htbp]
	\centering
	\includegraphics[width=0.55\hsize,bb= 0 0 792 612]{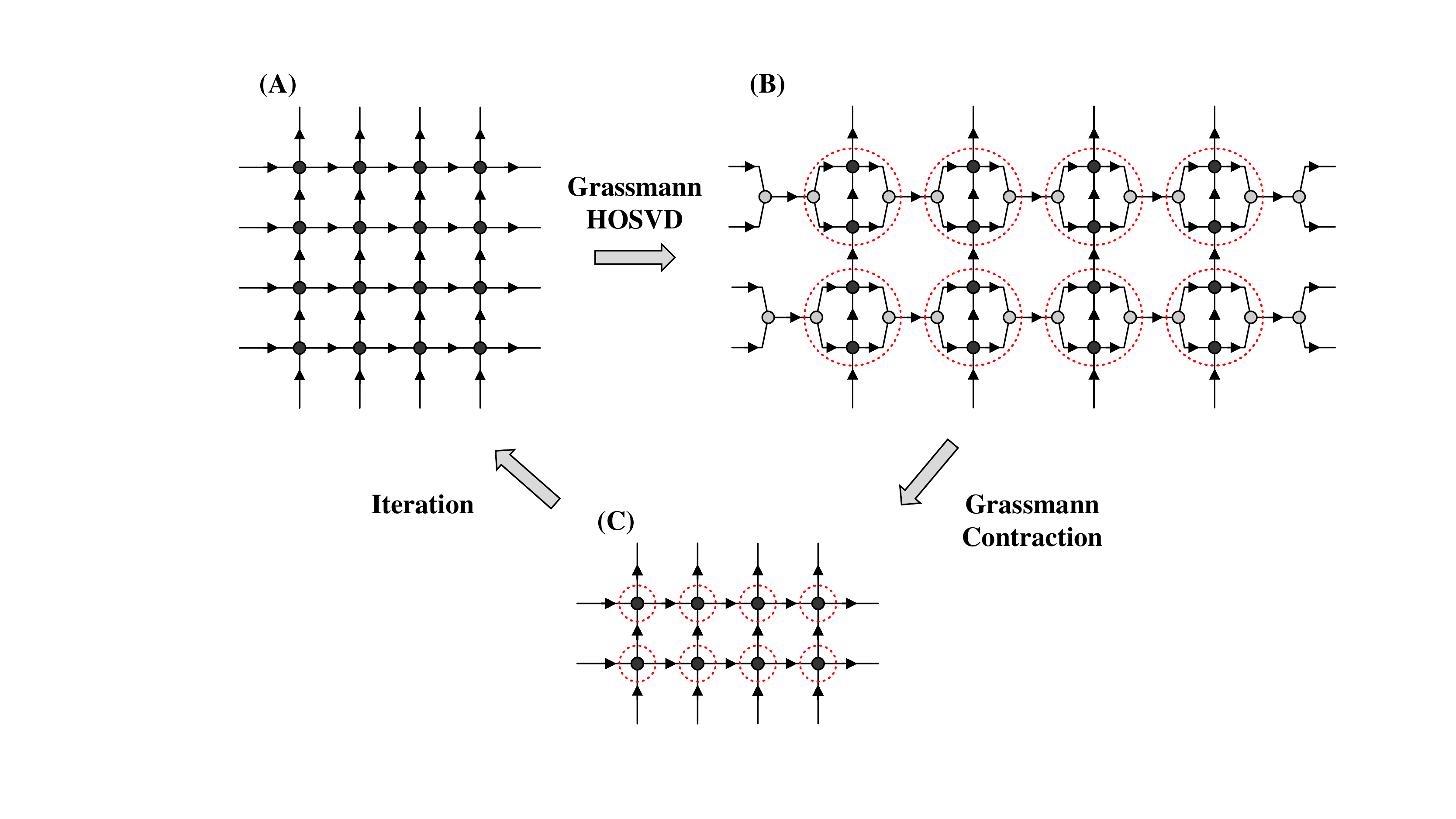}
	\caption{Schematic picture of the Grassmann HOTRG. 
   (A) Grassmann tensor network in two dimensions. 
   (B) Grassmann isometries are inserted in the whole network. (C) Tensor network is renormalized so that the lattice size is reduced by a factor of 2.}
	\label{fig:ghotrg}
\end{figure}

Let us consider a two-dimensional case as an example. 
The Grassmann tensor network is made of a $4K$-rank Grassmann tensor, which is identified as the Grassmann tensor $\mathcal{T}_{XY\bar{Y} \bar{X}}$ 
with four $K$-component indices $X,\bar{X},Y,\bar{Y}$.
Here, $K$ is the number of hopping terms.
Hereafter we count the rank of a Grassmann tensor in terms of $K$-component index. 
We assume that $X$ and $Y$ live on the links $(n,n+\hat{\mu})$ for $\mu=1,2$, respectively and $\bar{X}$ and $\bar{Y}$ live on the links $(n,n-\hat{\mu})$ for $\mu=1,2$, respectively.

Fig.~\ref{fig:ghotrg} schematically illustrates the algorithm of the {\it Grassmann HOTRG}, 
which employs the higher-order singular value decomposition (HOSVD) for the coefficient tensor of
\begin{align}
\label{eq:M_HOTRG}
	\mathcal{M}_{X_{1}X_{2} Y \bar{Y}\bar{X}_{2}\bar{X}_{1}} = 
\int\diff\bar{\Xi}\diff\Xi \ 
\mathcal{T}_{X_{2}\Xi\bar{Y}\bar{X}_{2}} \mathcal{T}_{X_{1}Y\bar{\Xi}\bar{X}_{1}}.
\end{align}
$\mathcal{M}$ is identified as a Grassmann tensor of rank $6$, 
and the coefficient tensor $M$ which is read from Eq.~(\ref{eq:M_HOTRG}) 
is given by a contraction of coefficient tensor $T$ with some sign factors.
We can decompose $\mathcal{M}$ in a formal way,
\begin{align}
\label{eq:ghosvd}
	\mathcal{M}_{X_{1}X_{2}Y\bar{Y}\bar{X}_{2}\bar{X}_{1}}
=\left(\prod_{k=1}^{4}\int\diff\bar{\Xi}_{k}\diff\Xi_{k}
\right) 
\mathcal{U}^{A}_{X_{1}X_{2}\Xi_{1}}\mathcal{U}^{B}_{Y\Xi_{2}}\mathcal{U}^{C}_{\bar{Y}\Xi_{3}} 
\mathcal{U}^{D}_{\bar{X}_{2}\bar{X}_{1}\Xi_{4}}\mathcal{S}_{\bar{\Xi}_{4}\bar{\Xi}_{3}\bar{\Xi}_{2}\bar{\Xi}_{1}}.
\end{align}
This decomposition is referred to as the {\it Grassmann HOSVD}, which is equivalent to the HOSVD 
for the coefficient tensor $M$. The Grassmann HOSVD gives us a {\it Grassmann isometry},
\begin{align}
\label{eq:isometry}
	\int\diff\bar{\Phi}\diff\Phi \ 
	\mathcal{U}_{\bar{X}_{2}\bar{X}_{1}\Phi}\mathcal{U}^{*}_{\bar{\Phi} X_{1} X_{2}},
\end{align}
which is inserted into the Grassmann tensor network to truncate the bond degrees of 
freedom (Fig.~\ref{fig:ghotrg}(B)). $\mathcal{U}$ is chosen from $\mathcal{U}^{A}$ and $\mathcal{U}^{D}$ in 
Eq.~\eqref{eq:ghosvd}, following the algorithm of the HOTRG \cite{PhysRevB.86.045139}. 
Note  that the dimension of $\Phi$ ($\bar \Phi$) is originally the square of
that of $X_m$ because $\mathcal{U}_{\bar X_2\bar X_1\Phi}$ is a square matrix with the column $X_1,X_2$
and the row $\Phi$. 
For a given bond dimension $D$, we truncate $\mathcal{U}$  so that the effective dimension of coefficient tensor associated with $\Phi$ ($\bar \Phi$) runs up to $D$, setting extra elements of $\mathcal{U}$ to zero. 
\footnote{
This can be formally achieved as follows: 
let $k$ be an integer such that $2^{k}$ is the least integer greater than 
or equal to $D$.  
We set the elements from column $D+1$ to column $2^{k}$ of coefficient tensor in $\mathcal{U}$  to zero 
to make the bond dimension of renormalized tensor become $D$ effectively.
}
However, a decimal numeral system defined in
 Ref.~\cite{Kadoh} (or see Appendix~\ref{sec:truncation}) allows us just 
to pick up $D^{2}\times D$ elements in the coefficient tensor in $\mathcal{U}$. 
This is useful to implement the current Grassmann HOTRG in practice.

Renormalized Grassmann tensor is finally defined by 
\begin{align}
	\mathcal{T}^{(1)}_{X Y\bar{Y}\bar{X}}=\left(\prod_{i=1}^{2}
	\int\diff\bar{Z}_{i} \diff{Z_{i}}  
       \int\diff\bar{Z}'_{i}\diff{Z'_{i}} 
       \right)
\mathcal{U}_{\bar{Z}_{2}\bar{Z}_{1}X}
\mathcal{M}_{Z_{1}Z_{2}Y\bar{Y}\bar{Z}'_{2}\bar{Z}'_{1}}\mathcal{U}_{\bar{X}Z'_{1}Z'_{2}}.
\end{align}
Repeating the above procedure, $Z$ can be approximated by 
\begin{align}
\label{eq:periodic}
	Z \approx \int\diff\bar{X}\diff X
	\int\diff\bar{Y}\diff Y \ 
	\mathcal{T}^{(n)}_{X Y\bar{Y}\bar{X}}.
\end{align}
Here, we assume that the lattice theory is defined on a finite lattice of $V=2^n$ with 
the periodic boundary condition \footnote{If one imposes the anti-periodic boundary condition in $2$-direction, all the Grassmann numbers in $Y$ or $\bar{Y}$ should be multiplied by $-1$ before carrying out the integration in Eq.~\eqref{eq:periodic}.} 
and $\mathcal{T}^{(n)}$ is the renormalized tensor at $n$th renormalization step.

It is worth emphasizing that in the above procedure, the original Grassmann tensor $\mathcal{T}_{XY\bar{Y}\bar{X}}$ is converted into the coarse-grained one 
$\mathcal{T}^{(1)}_{XY\bar{Y}\bar{X}}$. Since both of them are defined via Eq.~\eqref{eq:def1}, the current formulation recursively introduces the Grassmann tensor under the TRG procedure. Thanks to this property, we can avoid any ad-hoc treatment in defining the initial tensor network and coarse-grained one as in Ref.~\cite{Sakai:2017jwp}.

We examine the above Grassmann HOTRG by benchmarking with the one-flavor colorless free 
Wilson fermion and free staggered fermions on a two-dimensional square lattice. 
Unless otherwise noted, we assume the anti-periodic boundary condition in $2$-direction. 
Initial tensors for these fermions are obtained via Eq.~\eqref{eq:Gtensor}. See Appendices \ref{sec:wilson} and \ref{sec:staggered} for their concrete forms.

\begin{figure}[htbp]
	\centering
	\includegraphics[width=0.6\hsize,bb= 0 0 792 612]{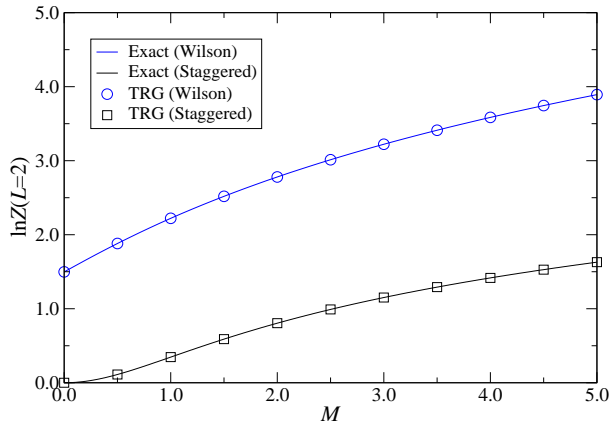}
	\caption{Free energy densities for free Wilson and staggered fermions against the mass $M$ on $2\times2$ lattice with $D=16$. They agree with the exact values up to the machine precision. }
	\label{fig:l2}
\end{figure}

\begin{figure}[htbp]
	\centering
	\includegraphics[width=0.6\hsize,bb= 0 0 792 612]{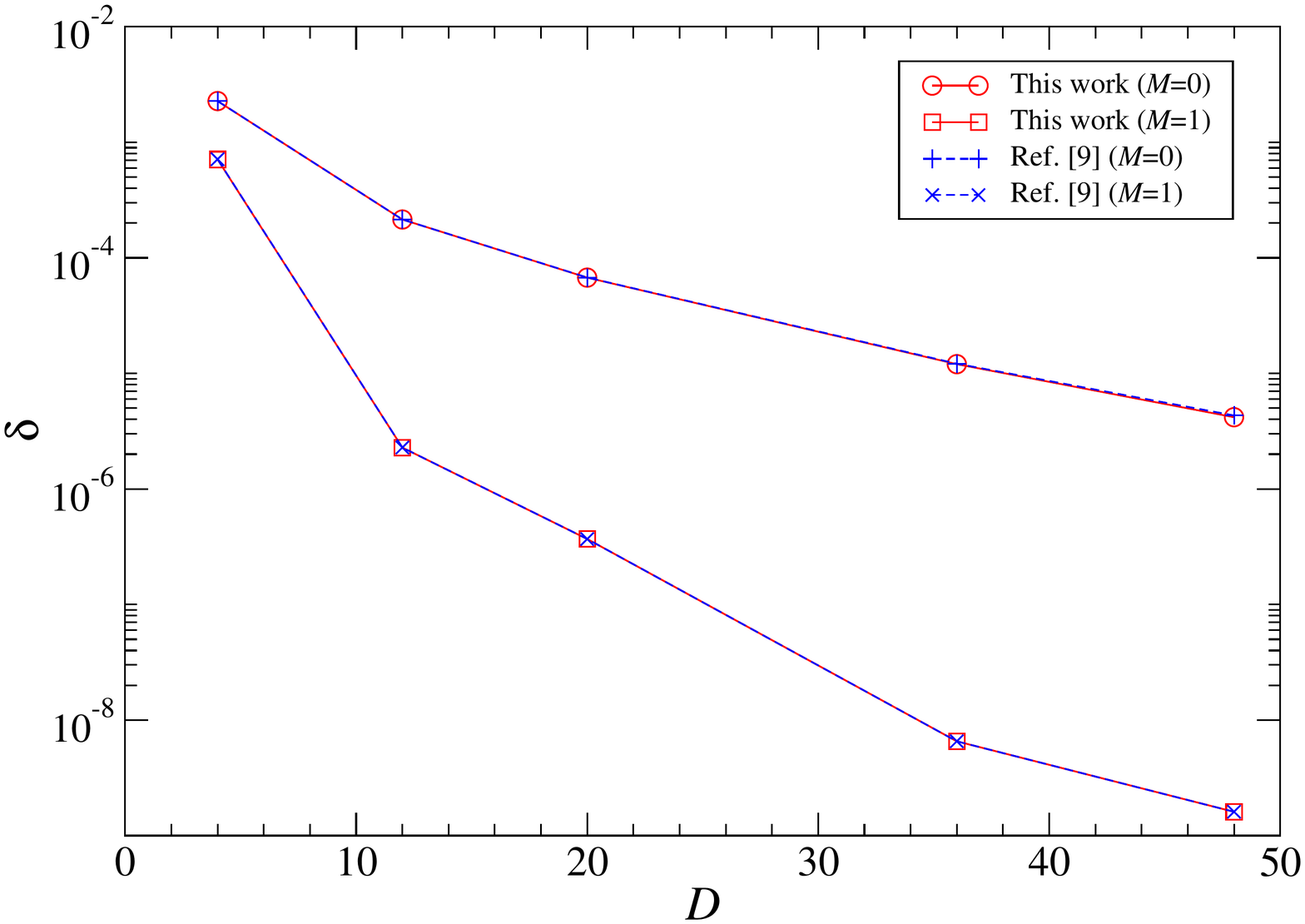}
	\caption{Relative error for the free energy of Wilson fermions on $1024\times1024$ lattice as a function of $D$. }
	\label{fig:delta}
\end{figure}

Fig.~\ref{fig:l2} shows the free energy per site against the mass $M$ on $2\times2$ 
lattice with $D=16$. With the choice of $D\ge16$, the calculation by the current 
Grassmann HOTRG agrees with the exact results up to the machine precision. 
\footnote{This situation is completely the same with the conventional Grassmann HOTRG~\cite{Sakai:2017jwp}.}
Fig.~\ref{fig:delta} plots the relative error for the free energy of free Wilson fermions on $1024\times1024$ lattice, defined by
\begin{align}
\label{eq:delta}
	\delta=\left|\frac{\ln Z(L=1024,D)-\ln Z_{\mathrm{exact}}(L=1024)}{\ln Z_{\mathrm{exact}}(L=1024)}\right|.
\end{align}
It is confirmed that the current Grassmann HOTRG has achieved the same accuracy as the conventional one \cite{Sakai:2017jwp} both for massless and massive fermions.

\begin{figure}[htbp]
	\centering
	\includegraphics[width=0.6\hsize,bb= 0 0 792 612]{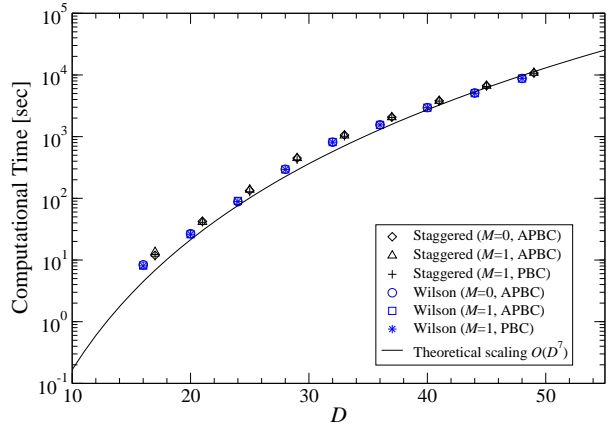}
	\caption{Computational time of free energy density on $1024\times1024$ lattice as a function of $D$. For the massive cases, we evaluated the path integrals assuming the periodic boundary condition (PBC) and anti-periodic one (APBC) for $2$-direction. Solid curve shows the theoretical scaling of computational time, which holds whether the fermions are massless or massive regardless of the type of lattice fermion.}
	\label{fig:time_d}
\end{figure}

Fig.~\ref{fig:time_d} shows the computational time as a function of the bond dimension for the Wilson or staggered fermions with various conditions. The solid curve represents the theoretical scaling of the computational time of the HOTRG, which is $O(D^{7})$ in 2 dimensions, and the current Grassmann HOTRG computation well reproduces it. 

\begin{figure}[htbp]
	\centering
	\includegraphics[width=0.6\hsize,bb= 0 0 792 612]{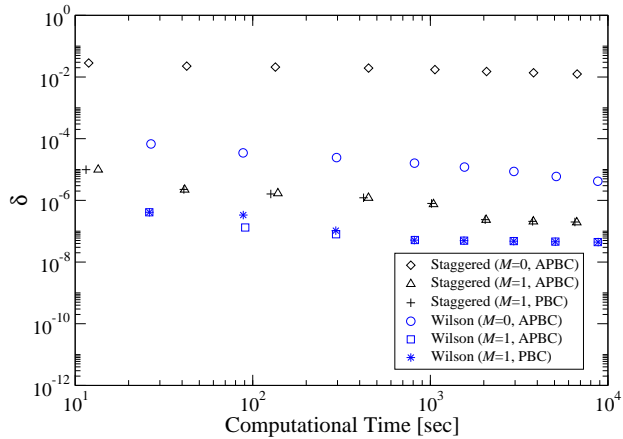}
	\caption{Relative error of free energy on $1024\times1024$ lattice as a function of the computational time. Different symbol corresponds to the Wilson or staggered fermions with $M=0, 1$. For the massive cases, we evaluated the path integrals assuming the periodic boundary condition (PBC) and anti-periodic one (APBC) for $2$-direction.}
	\label{fig:delta_time}
\end{figure}

Finally, we evaluate the relative error defined via Eq.~\eqref{eq:delta} as a function of the computational time for both Wilson and staggered fermions, varying values of mass, and boundary conditions. As shown in Fig.~\ref{fig:delta_time}, with the vanishing mass, the Grassmann HOTRG achieves the higher accuracy for the Wilson fermions compared to staggered fermions with the fixed computational time. 
This may be attributed to the chiral symmetry in staggered fermions, which makes the hierarchy of the singular 
values in the Grassmann tensor milder. 
When these fermions are massive, the Grassmann HOTRG reaches slightly higher accuracy for Wilson fermions within the fixed computational time. We should note that the situation is quite different from the Monte Carlo simulation, where the computational time of staggered fermions is faster than that of the other lattice fermions. Fig.~\ref{fig:delta_time} suggests that Wilson fermions show better performance than staggered fermions in the TRG method with the fixed execution time, unlike the Monte Carlo method.

\section{Summary and outlook}
\label{sec:summary}

A tensor network formulation for fermion theories was discussed, based on the introduction of the auxiliary fermion fields. 
We derived a general formula of  the tensor network representation for lattice fermions. 
This formula is immediately applicable for many types of local lattice fermions such as Wilson fermions,
staggered fermions and domain-wall fermions. 
Our method is useful in practice because it allows us to recursively introduce the coarse-grained Grassmann tensor 
throughout the TRG calculation and provides a fair comparison among various lattice fermions from a common aspect.
Implementing the Grassmann HOTRG, whose accuracy are the same as in Ref.~\cite{Sakai:2017jwp}, our numerical results suggest that Wilson fermions show better performance than staggered fermions in the TRG method with the fixed execution time. 
The situation is quite different from the Monte Carlo method, and
these results would be interesting as a starting point of further studies, not only for free field theories but also interacting ones.

\begin{figure}[htbp]
	\centering
	\includegraphics[width=0.75\hsize,bb= 0 0 792 612]{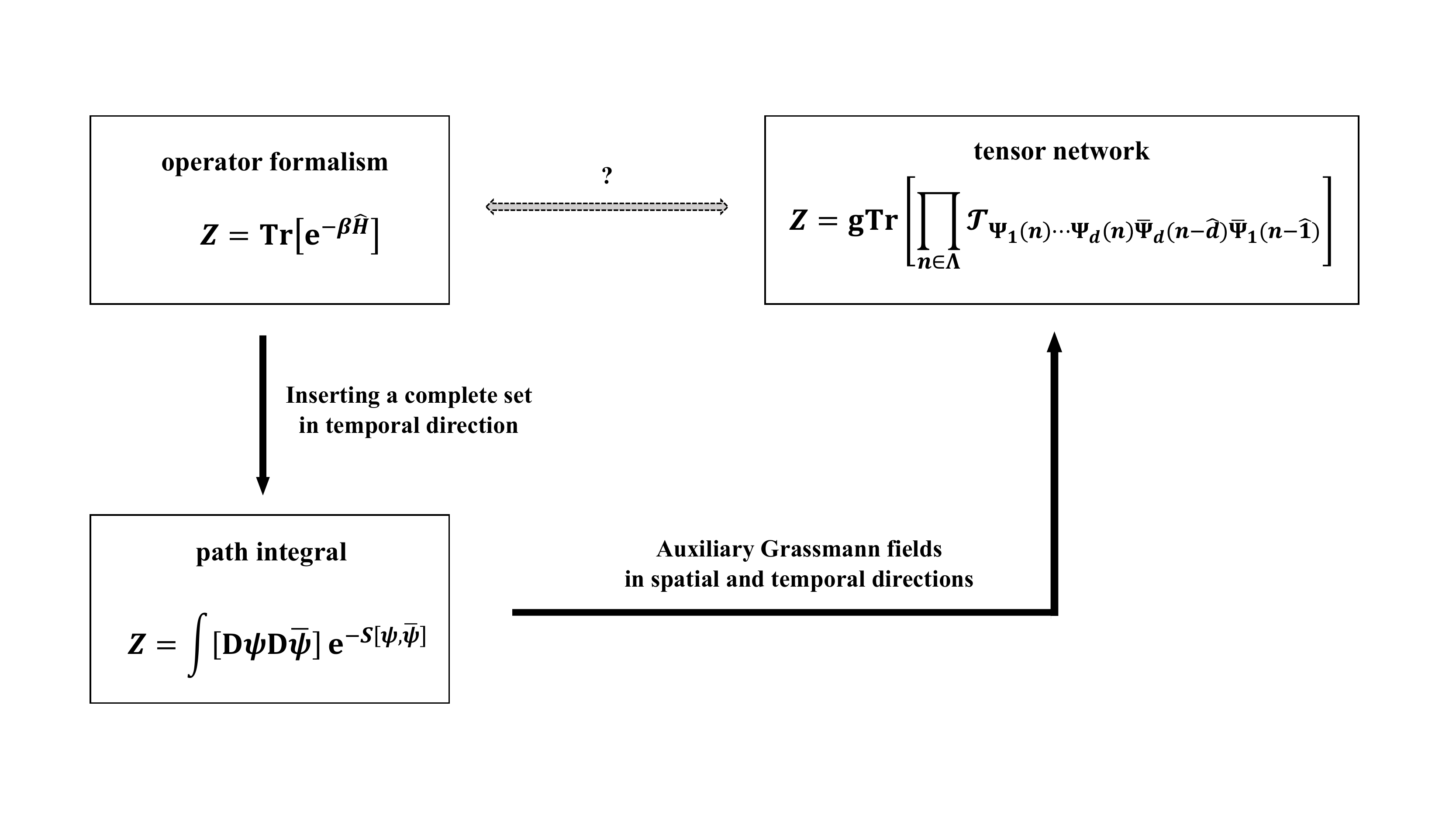}
	\caption{A possible relationship between the path integral, operator formalism, and tensor network.}
	\label{fig:commutative}
\end{figure}

It is worth noting that the current formulation depends on the introduction of auxiliary 
Grassmann fields both in spatial and temporal directions for the path integral $Z$. 
On the other hand, the path integral is derived from $Z={\rm Tr}({\rm e}^{-\beta \hat H})$ 
inserting a complete set of coherent fermion states. 
This implies that a tensor network representation for lattice fermions could be derived directly from 
$Z={\rm Tr} ({\rm e}^{-\beta \hat H})$.  Fig.~\ref{fig:commutative} shows a possible relationship between the path integral, operator formalism, and tensor network. This viewpoint may be useful in extending our method to interacting theories with gauge fields.

\appendix

\input{app1}

\input{app2}

\acknowledgments

We are grateful to Yoshinobu Kuramashi, Ryo Sakai, Shinji Takeda, Yusuke Yoshimura for their insightful discussions.
This work is supported by the JSPS KAKENHI Grant JP19K03853, JP21J11226.

\bibliographystyle{JHEP}
\bibliography{algorithm,continuous,discrete,formulation,gauge,grassmann,gravity}

\end{document}

%% file: app1.tex
\section{Truncation technique}
\label{sec:truncation}

We introduce the SVD of a Grassmann tensor, 
which is equivalent to the SVD for the corresponding coefficient tensor. 
Let $\mathcal{T}_{\Psi\Phi}$ be a Grassmann tensor whose rank is $2N$. 
We represent the coefficient tensor of $\mathcal{T}_{\Psi\Phi}$ as $2^{N}\times2^{N}$ 
matrix $T_{IJ}$ with $I=(i_{1},\cdots,i_{N})$ and $J=(i_{N+1},\cdots,i_{2N})$. 
Since the Grassmann parity of $\mathcal{T}_{\psi\phi}$ is even, $T_{IJ}$ takes a non-zero value if and only if
\begin{align}
	\sum_{k=1}^{2N}i_{k}\mod2=0
\end{align}
is satisfied. This condition allows us to obtain a block diagonal matrix representation for $T_{IJ}$. According to Ref.~\cite{Kadoh}, we now define the following decimal numeral system,
\begin{align}
	I=
	\begin{cases}
		\sum_{k=1}^{2N}2^{k-1}i_{k}~~~&(i_{2}+\cdots+i_{2N}\mod2=0) \\
		1-i_{1}+\sum_{k=2}^{2N}2^{k-1}i_{k}~~~&(i_{2}+\cdots+i_{2N}\mod2=1).
	\end{cases}
\end{align}
Thanks to this decimal numeral system, the parity of $\sum_{k=1}^{2N}i_{k}$ corresponds with that of $I$. Applying this system for $I$ and $J$ in $T_{IJ}$, one obtains the block diagonal matrix
\begin{align}
	T=
	\begin{bmatrix}
		T^{\mathrm{E}} & 0 \\
		0 & T^{\mathrm{O}}
	\end{bmatrix}
	.
\end{align} 
The SVD for $T$ is obtained from that for $T^{\mathrm{E}}$ and $T^{\mathrm{O}}$,
\begin{align}
\label{eq:even}
	T^{\mathrm{E}}_{IJ}=\sum_{K:\mathrm{even}}U^{\mathrm{E}}_{IK}\sigma^{\mathrm{E}}_{K}V^{\mathrm{E}}_{JK},
\end{align}
\begin{align}
\label{eq:odd}
	T^{\mathrm{O}}_{IJ}=\sum_{K:\mathrm{odd}}U^{\mathrm{O}}_{IK}\sigma^{\mathrm{O}}_{K}V^{\mathrm{O}}_{JK}.
\end{align}
Picking up the largest $D$ numbers of singular values and corresponding singular vectors, $T$ is approximated with a lower-rank matrix. 

We apply the above technique for $MM^{\dag}$, where $M$ is the coefficient tensor in Eq.~\eqref{eq:M_HOTRG}. In Eq.~\eqref{eq:isometry}, the Grassmann isometry defines new bond Grassmann numbers $\Phi$ and $\bar{\Phi}$. It is worth emphasizing that the parity of these Grassmann numbers corresponds to the parity of $K$ in Eqs.~\eqref{eq:even} and \eqref{eq:odd}. This property is significantly useful in developing the current Grassmann HOTRG.

%% file: app2.tex
\section{Tensor networks for 2D lattice fermions}
\label{sec:2DTN}

\subsection{Wilson fermions}
\label{sec:wilson}

The Dirac matrix with the Wilson parameter $r=1$ is given by
\begin{align}
	D(n,m)&=(M+2)
\begin{bmatrix}
	\delta(n,m) & 0 \\
	0 & \delta(n,m)
\end{bmatrix}
\nonumber\\
&\quad-\frac{1}{2}
\begin{bmatrix}
	\ \delta(n-\hat{x},m)+\delta(n+\hat{x},m) \ & \ \delta(n-\hat{x},m)-\delta(n+\hat{x},m) \ \\
	\ \delta(n-\hat{x},m)-\delta(n+\hat{x},m) \ & \ \delta(n-\hat{x},m)+\delta(n+\hat{x},m) \
\end{bmatrix}
\nonumber\\
&\quad-
\begin{bmatrix}
	\delta(n-\hat{t},m) & 0 \\
	0 & \delta(n+\hat{t},m)
\end{bmatrix}
.
\end{align}
Applying the SVD, we have
\begin{align}
	D(n,m)&=(M+2)
\begin{bmatrix}
	\delta(n,m) & 0 \\
	0 & \delta(n,m)
\end{bmatrix}
+U_{x}
\begin{bmatrix}
	\delta(n-\hat{x},m) & 0 \\
	0 & \delta(n+\hat{x},m)
\end{bmatrix}
V_{x}^{\dagger}\nonumber\\
&\quad+U_{t}
\begin{bmatrix}
	\delta(n-\hat{t},m) & 0 \\
	0 & \delta(n+\hat{t},m)
\end{bmatrix}
V_{t}^{\dagger}
,
\end{align}
where
\begin{align}
	U_{x}=\frac{1}{\sqrt{2}}
\begin{bmatrix}
	-1 & -1 \\
	-1 & 1
\end{bmatrix}
,~~
	U_{t}=
\begin{bmatrix}
	-1 & 0 \\
	0 & -1
\end{bmatrix}
,
\end{align}
and $V_{\mu}^{\dagger}=-U_{\mu}$ for $\mu=x,t$. Following Eq.~\eqref{eq:Gtensor}, the Grassmann tensor $\mathcal{T}_{\Psi_x\Psi_t\bar{\Psi}_t\bar{\Psi}_x}$ with the ordering $\Psi_\mu=(\eta_{\mu},\zeta_{\mu})$ and $\bar{\Psi}_{\mu}=(\bar{\zeta}_{\mu},\bar{\eta}_{\mu})$ for $\mu=x,t$ is obtained immediately. One finds
\begin{align}
\label{eq:explicit}
	&\hspace{-5mm} \mathcal{T}_{\Psi_x\Psi_t\bar{\Psi}_t\bar{\Psi}_x} %\nonumber\\
	=(M+2)^2 -D_{22} B_{11} B_{11}\eta_{t}\bar{\eta}_{t}
	-D_{11} B_{22} B_{22}\zeta_{t}\bar{\zeta}_{t} 
	\nonumber \\
	%%% diagonal
	&
	-\left[
	D_{11} A_{21} A_{12}
	+D_{22} A_{11} A_{11}
	\right]\eta_{x}\bar{\eta}_{x} 
	-\left[
	D_{11} A_{22} A_{22}
	+D_{22}  A_{12} A_{21}
	\right]\zeta_{x}\bar{\zeta}_{x}\nonumber\\
	%
	%
	%&-D_{22} B_{11} B_{11}\eta_{t}\bar{\eta}_{t}
	%-D_{11} B_{22} B_{22}\zeta_{t}\bar{\zeta}_{t} \nonumber\\
	%
	%
	%%% vanishing
	&
	+\left[
	D_{11} A_{22} A_{12} + D_{22} A_{12} A_{11}
	\right]\eta_{x}\zeta_{x} %\nonumber\\
	%
	%
	%&\quad 
	- \left[
	D_{11} A_{21} A_{22}+D_{22} A_{11} A_{21}
	\right]\bar{\zeta}_{x}\bar{\eta}_{x}\nonumber\\
	&+D_{11} B_{22} A_{12}\eta_{x}\zeta_{t}
	- D_{11} A_{21} B_{22}\bar{\zeta}_{t}\bar{\eta}_{x} %\nonumber\\
	%
	%
	%&\quad 
	-D_{11} A_{22} B_{22}\zeta_{x}\bar{\zeta}_{t}
	-D_{11} B_{22} A_{22}\zeta_{t}\bar{\zeta}_{x}\nonumber\\
	& -D_{22} A_{12} B_{11}\zeta_{x}\eta_{t}
	+D_{22} B_{11} A_{21}\bar{\eta}_{t}\bar{\zeta}_{x} %\nonumber\\
	%
	%
	%&\quad 
	-D_{22} B_{11} A_{11}\eta_{x}\bar{\eta}_{t}
	-D_{22} A_{11} B_{11}\eta_{t}\bar{\eta}_{x}\nonumber\\
	&-\left[
	\zeta_{t}\bar{\eta}_{t}-\zeta_{x}\bar{\eta}_{x}
	+A_{11} B_{22}\zeta_{t}\bar{\eta}_{x}
	+B_{11} A_{21}\bar{\eta}_{t}\bar{\eta}_{x} %\right.\nonumber\\
	%&\quad\left.\quad\quad\quad\quad\quad\quad\quad
	+A_{12} B_{22}\zeta_{x}\zeta_{t}
	+B_{11} A_{22}\zeta_{x}\bar{\eta}_{t}
	\right]\nonumber\\
	&\quad\times\left[
	\eta_{t}\bar{\zeta}_{t}-\eta_{x}\bar{\zeta}_{x}
	+ B_{22} A_{11}\eta_{x}\bar{\zeta}_{t}
	-A_{12} B_{11}\eta_{x}\eta_{t} %\right.\nonumber\\
	%&\quad\left.\quad\quad\quad\quad\quad\quad\quad
	-B_{22} A_{21}\bar{\zeta}_{t}\bar{\zeta}_{x}
	+A_{22} B_{11}\eta_{t}\bar{\zeta}_{x}
	\right], 
\end{align}
where $D_{ab}\equiv D_{ab}(n,n)$,  $A=U_x,B=U_t$.

\subsection{Staggered fermions}
\label{sec:staggered}

The action of the two-dimensional staggered fermions is given by
\begin{align}
	S=\sum_{n=(n_{x},n_{t})\in\Lambda}\bar{\chi}(n)\left[\sum_{\mu=x,t}p_{\mu}(n)\frac{\chi(n+\hat{\mu})-\chi(n-\hat{\mu})}{2}+M\chi(n)\right],
\end{align}
where $\chi(n)$ and $\bar{\chi}(n)$ are single-component Grassmann fields and $p_{\mu}(n)$ is the staggered sign function defined by $p_{x}(n)=1$ and $p_{t}(n)=(-1)^{n_{x}}$.
Since there is no Dirac structure in the action of the staggered fermion, we can immediately derive the tensor network representation for the path integral without applying SVD. Employing Eqs.~\eqref{eq:basic_id1} and \eqref{eq:basic_id2}, we can obtain
\begin{align}
	\mathcal{T}_{\Psi_x(n)\Psi_t(n)\bar{\Psi}_t(n-\hat{t})\bar{\Psi}_x(n-\hat{x})}
	= &-M
	-\frac{1}{2} \eta_{x}\bar{\eta}_{x}
	+\frac{1}{2} \zeta_{x}\bar{\zeta}_{x}
	-\frac{p_{t}(n)}{2} \eta_{t}\bar{\eta}_{t}
	+\frac{p_{t}(n)}{2} \zeta_{t}\bar{\zeta}_{t}
	\nonumber\\
	&-\frac{1}{2} \eta_{x}\zeta_{x}
	-\frac{1}{2} \eta_{t}\zeta_{t}
	-\frac{1}{2} \bar{\zeta}_{x}\bar{\eta}_{x}
	-\frac{1}{2} \bar{\zeta}_{t}\bar{\eta}_{t}
	\nonumber\\
	&-\frac{p_{t}(n)}{2} \eta_{x}\zeta_{t}
	+\frac{p_{t}(n)}{2} \zeta_{x}\eta_{t}
	+\frac{1}{2} \bar{\zeta}_{t}\bar{\eta}_{x}
	-\frac{1}{2} \bar{\eta}_{t}\bar{\zeta}_{x}
	\nonumber\\
	&-\frac{1}{2} \eta_{x}\bar{\eta}_{t}
	+\frac{1}{2} \zeta_{x}\bar{\zeta}_{t}
	-\frac{p_{t}(n)}{2} \eta_{t}\bar{\eta}_{x}
	+\frac{p_{t}(n)}{2} \zeta_{t}\bar{\zeta}_{x},
\end{align}
with $\Psi_\mu=(\eta_{\mu},\zeta_{\mu})$ and $\bar{\Psi}_{\mu}=(\bar{\zeta}_{\mu},\bar{\eta}_{\mu})$ for $\mu=x,t$. Note that the resulting Grassmann tensor depends on the parity of $n_{x}$.